\documentclass{elsart}

\usepackage{amssymb,amsfonts,graphicx}
\usepackage{dcolumn}
\usepackage{bm}

\begin{document}

\begin{frontmatter}

\title{Group dynamics of the Japanese market}

\author[bu,kaist]{Woo-Sung Jung}
\ead{wsjung@physics.bu.edu}
\author[ku]{Okyu Kwon}
\author[bu]{Fengzhong Wang}
\author[icu]{Taisei Kaizoji}
\author[kaist]{Hie-Tae Moon}
\author[bu]{H. Eugene Stanley}

\address[bu]{Center for Polymer Studies and Department of Physics, Boston University, Boston, MA 02215, USA}
\address[kaist]{Center for Complex Systems and Department of Physics, Korea Advanced Institute of Science and Technology, Deajeon 305-701, Republic of Korea}
\address[ku]{Department of Physics, Korea University, Seoul 136-701, Republic of Korea}
\address[icu]{Division of Social Sciences, International Christian University, Tokyo 181-8585, Japan}

\begin{abstract}

We investigated the network structures of the Japanese
stock market through the minimum spanning tree. We defined
\textit{grouping coefficient} to test the validity of conventional
grouping by industrial categories, and found a decreasing in trend
for the coefficient. This phenomenon supports the increasing
external influences on the market due to the globalization. To
reduce this influence, we used S\&P500 index as the international
market and removed its correlation with every stock. We found
stronger grouping in this measurement, compared to the original
analysis, which agrees with our assumption that the international
market influences to the Japanese market.

\end{abstract}


\end{frontmatter}

\section{Introduction}
The network theory is generally used to investigate complex systems
with many interacting agents. The financial market, where all listed
companies are correlated with each other, has received attention as
a typical complex system \cite{Mantegna1999B,Arthur1997B}. A popular
method is the minimum spanning tree (MST) which constructs the asset
tree using correlations between stock prices
\cite{Mantegna1999,Micciche2003,Onnela2004,Coronnello2005,Bonanno2003,Mizuno2006,Coelho2006,Toth2006}.
It gives us the characteristic feature of the market in the simple
way. In a MST of $N$ nodes, each node represents a company and $N-1$
links with the most important correlations are selected. The MST is
a loop-less network, and every node at least has a link. Also, the
grouping of companies in the MST can be identified and extended to
portfolio optimization and the companies of the US market are
clearly grouped with the industry category or business sector
\cite{Onnela2003}.

Several papers show that the characteristics of the mature market cannot be simply extended to the emerging market \cite{Matia2004,Yan2005} and change dynamically due to the globalization \cite{Oh2007}.
Especially, the Korean market is synchronized to external markets
due to the globalization, and the tendency of grouping by industry
categories disappeared \cite{Jung2006B}. In addition, the Korean
market forms clusters according to the MSCI index after 2000 and the
tendency of synchronization to the US market is stronger and
stronger \cite{Jung2006}.

The Japanese market attracts many econophysicists since its unique
character \cite{Ishikawa2006,Kaizoji2000}. It is not an emerging one
but it has a strong connection with the Asian emerging markets. For
instance, the Korean market has many common features with the
Japanese market. Korea and Japan have developed close
interdependence of their economic systems for a long time. Two
countries experienced very high rate of economic growth. Their
economies were commonly driven by the government-directed investment
model and the protective trade policy at first. In addition, the
integration into the global economy has much developed in the recent
days, especially in the Asia. Such violent changes make the studies
of the Asian financial markets very interesting.

We investigated characteristics of the Japanese stock market with
the history of the market. First, we compared the analysis of the
Japanese market with that of the Korean market and found common
features of these two markets such as decreasing of grouping by
industry categories. Second, we used S\&P500 index as the
international market to reduce external influences due to the
globalization. This study showed that the international market
obviously influences the Japanese market.

\section{Characteristics of the Japanese market}
\begin{figure}
\resizebox{1.00\columnwidth}{!}{
\includegraphics{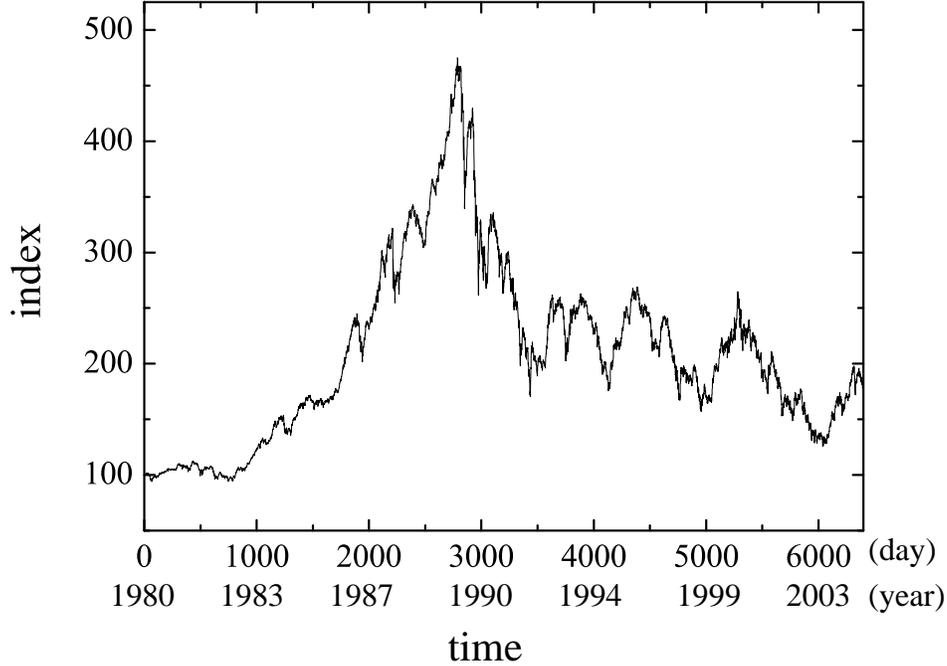}
}
\caption{The custom-made index for 624 Japanese companies selected in the Tokyo Stock Exchange from 1980 to 2004.} \label{indexjapan}
\end{figure}

There are several stock exchange markets in Japan, and we selected
the Tokyo Stock Exchange (TSE), the largest Japanese market. We used
the daily closing stock price from 1980 to 2004, and a total of 624
companies remained for the full 25 year-period were selected for our
analysis. To demonstrate the trend of the Japanese market, we plot
the index which averages over the 624 companies in Fig.
\ref{indexjapan}.

The logarithmic return is defined as $S_i(t)\equiv\ln J_i(t+\Delta
t) - \ln J_i (t)$, where $J_i(t)$ is the closing price of a given
Japanese company $i$. And, the cross-correlation coefficient between
stock $i$ and $j$ is defined as:

\begin{equation}
\rho_{ij} \equiv\frac{(<S_{i}S_{j}>-<S_{i}><S_{j}>)}
{\sqrt{(<S_{i}^{2}>-<S_{i}>^{2})(<S_{j}^{2}>-<S_{j}>^{2}) }}.
\label{correlation}
\end{equation}

To explore the evolution of the market, we construct the MSTs with
time windows of width $T$ corresponding to daily data for \emph{3
years}, sliding the window with 20 trading day, approximately 1
month. Each node of the MST corresponds to a company, and each link
has a weight $\rho_{ij}(=\rho_{ji})$, which is simply the value of
the cross-correlation coefficient.


To investigate the group dynamics of the financial market more
detail, we define a quantity, \textit{grouping coefficient}, to
measure how well groups with the industry categories,

\begin{equation}
G=\frac{n(\sum_{i\in C})}{n(\sum_{i})},
\end{equation}

\noindent where $n(\sum_{i})$ represents the number of whole links
in the network, and $n(\sum_{i\in C})$ is the number of links
between companies in the same industry category.

\begin{figure}
\resizebox{1.00\columnwidth}{!}{
\includegraphics{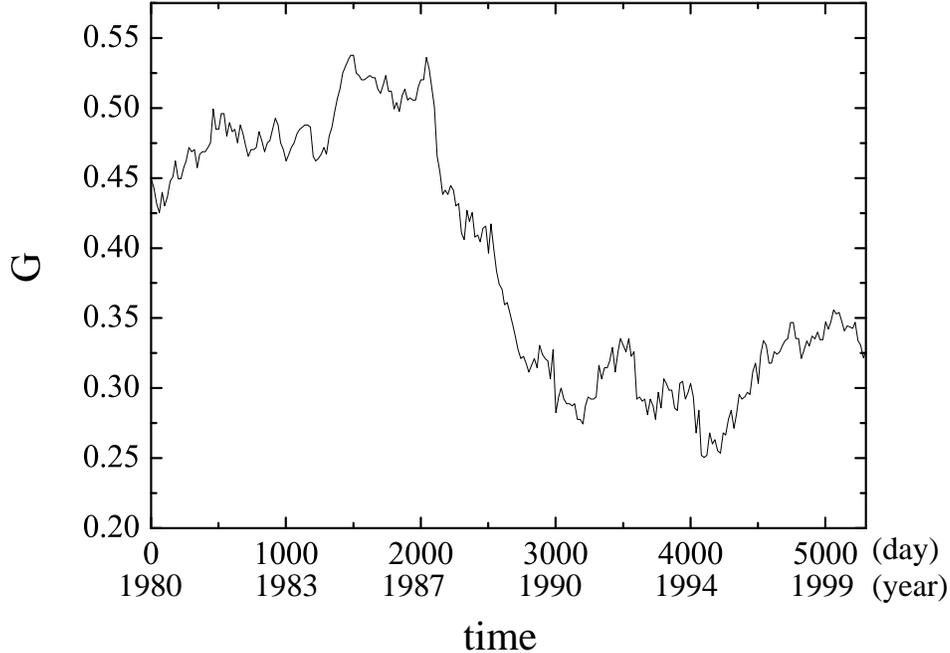}
}
\caption{Plot of the grouping coefficient for all categories as a
function of the starting time of the moving window from 1980 to
2004. }
\label{Gjapan}
\end{figure}

In our previous study \cite{Jung2006B}, we found the grouping
coefficient of the recent Korean market decreases with elapsing
time. In other words, the recent Korean market does not make the
groups by industry categories even though the recent US market's MST
is grouped well by industry categories \cite{Onnela2003}. As
expected, we observe the similar trend in the Japanese market (Fig.
\ref{Gjapan}). Before the mid-1980s, the coefficient $G$ of the
Japanese market shows no special movement. However, the coefficient
tends to decrease after the mid-1980s, and this tendency is
strengthened more and more by burst of the bubble in 1990's. It is
natural to form groups of industry categories, because companies
included in the same category are highly related to each other in
comparison to companies in other categories. However, the groups of
industry categories are breaking down in recent, and it is related
to the globalization of Asian markets. We will explain more in the
following section.

\section{Correlation with the US market}
\begin{figure*}
\resizebox{1.00\textwidth}{!}{
\includegraphics{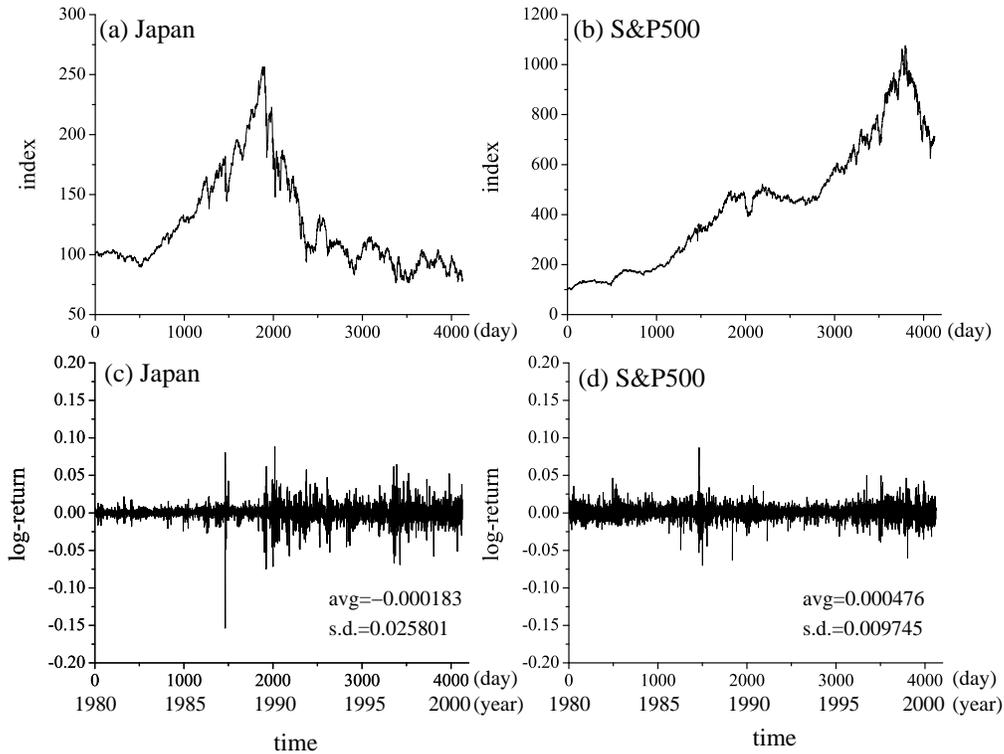}
}
\caption{The indexes and log returns of the Japanese and US markets for selected dates.} \label{indexjpus}
\end{figure*}

The US stock market is a dominating one over the whole world,
especially to the Asian markets. Thus, we choose the US market as a
representative of the international market in the section. Recently,
the US market's influence to Japan is more powerful in comparison
with the opposite direction
\cite{Climent2003,Eun1989,Becker1990,Karolyi1996}. We assume
$(n-1)$-th day's US market (represented by S\&P500 index) and $n$-th
day's Japanese market are highly correlated, and totally we have
4126 days from the 25-year data set. We use these days' closing data
of each Japanese company and S\&P500 index, and Fig. \ref{indexjpus}
shows the indexes and logarithmic returns for new data set of the
Japanese and US markets, respectively.

We define the \textit{modified logarithmic return} for a given
Japanese company $i$ at time $t$, $M_i(t)$, to minimize the
influence of the US market,

\begin{equation}
M_i(t) \equiv S_i(t)-\alpha\beta_i(t)U(t-1),
\label{modifiedreturn}
\end{equation}

\noindent where $S_i(t)$ represents a given Japanese company $i$'s
log return at time $t$ and $U(t-1)$ is the S\&P500 index's log
return at time $\left(t-1\right)$. The $\beta_i(t)$ value represents
the cross-correlation defined in Eq.(\ref{correlation}) between the
Japanese company $i$ and S\&P500 index within 3 years from time $t$,
and $\alpha$ is the rescaling coefficient.

\begin{figure}
\includegraphics{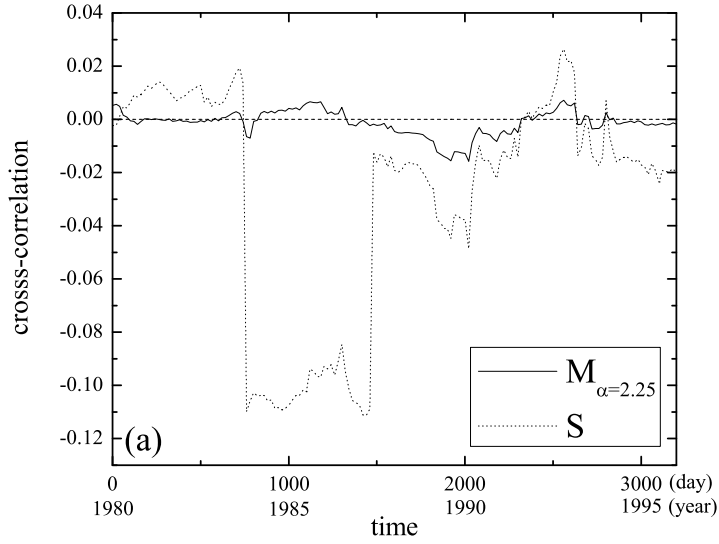}
\includegraphics{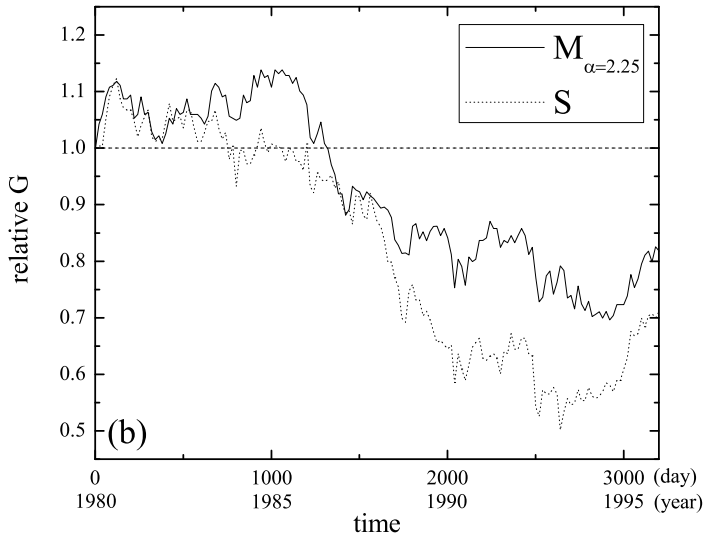}
\caption{(a) The cross-correlations between the Japanese companies' original logarithmic return, $S_i\left(t\right)$, or modified one, $M_i\left(t\right)$, and the S\&P500 index, $U\left(t-1\right)$. The dash line is a guideline of the value of 0.0.  (b) Rescaled grouping coefficients as a function of time for the Japanese market. They are calculated by $M_i(t)$ and $S_i(t)$, respectively. The dash line is a guideline of the value of 1.0.} \label{modifiedG}
\end{figure}

We calculate the cross-correlation between a Japanese company stock $i$'s logarithm return or modified one at time $t$ and the S\&P index at time $t-1$, and then take the average of these cross-correlations at the same time. Fig. \ref{modifiedG}(a) shows these averages as a function of the time (the dot line is for the original return while the solid line is for the modified one). If the US market's influence to the Japanese market is minimized, the correlation between the modified logarithm return of the Japanese market and the S\&P500 index approaches zero. Thus, we search the proper $\alpha$ value in Eq. \ref{modifiedreturn}, and find the value of 2.25 makes these cross-correlations almost zero. The meaning of the alpha value remains an open question. Other financial markets that synchronize to foreign market might have their own alpha value. 

According to Fig. \ref{modifiedG}(a), the US market and the original Japanese market are rather anti-correlated from 750-day to 1500-day. It is from a big crash of the Japanese market. In Fig. \ref{indexjpus}(a), it is found a crash at this period in the Japanese market. However, the index of the US market does not decrease (Fig. \ref{indexjpus}(b)), and the width of time windows $T$ in this paper is 3 $years$. Therefore, the cross-correlation between two markets has rather bigger negative value at this 3-year period which contains the Japanese market crash. That is why the sharp fall and jump are appeared in Fig. \ref{modifiedG}(a). It is interesting that the $alpha$ value of 2.25 also makes the cross-correlations zero through this anti-correlated period.

Fig. \ref{modifiedG}(b) represents the \textit{relative} grouping coefficients, $G(t)/G(t=0)$, calculated by $M_i(t)$ and $S_i(t)$, respectively. We use the relative values to compare two grouping coefficients' dynamical features together. Both of the grouping coefficients with $S(t)$ and $M(t)$ decrease after 1990. However, the tendency of the coefficient's decrease with $M(t)$ is smaller than that of $S(t)$. It means the decreasing of the grouping coefficient is related to the globalization. Nowadays, the Japanese market is synchronized to the US market \cite{Eun1989,Becker1990,Karolyi1996}, and the groups by the industry categories are broken down like the Korean market. In addition, the recent Korean market can be grouped in terms of the MSCI Korea Index \cite{Jung2006}. The companies included in the MSCI Korea index are more synchronized to a foreign market, and these companies make a group. The other groups consist of companies not included in the MSCI index. Thus, the grouping coefficient of the Korean market is decreasing in recent \cite{Jung2006B}.

\section{Conclusions}
We investigated the Japanese stock market networks using the daily closing stock price. We show that the grouping coefficients of the market decreased with elapsing time and the number of groups according to industry categories decreased. It is similar to the analysis of the Korean market, which has many common features with the Japanese one.

And, we defined the modified logarithmic return to minimize the external US market's influence. When it is minimized, the grouping coefficient is rather increased in comparison with the original coefficient. It means the decreasing of the coefficient is correlated to the external influences to the market. Currently, most world markets synchronize to the US market, and they are sensitive to foreign factors. The grouping coefficient could be a good measurement to quantitize this phenomenon.

\end{document}